\documentclass[twocolumn,english,aps,preprintnumbers,amsmath,amssymb,superscriptaddress]{revtex4}
\usepackage[latin9]{inputenc}
\usepackage{amsmath}
\usepackage{graphicx}

\makeatletter
\@ifundefined{textcolor}{}
{%
 \definecolor{BLACK}{gray}{0}
 \definecolor{WHITE}{gray}{1}
 \definecolor{RED}{rgb}{1,0,0}
 \definecolor{GREEN}{rgb}{0,1,0}
 \definecolor{BLUE}{rgb}{0,0,1}
 \definecolor{CYAN}{cmyk}{1,0,0,0}
 \definecolor{MAGENTA}{cmyk}{0,1,0,0}
 \definecolor{YELLOW}{cmyk}{0,0,1,0}
}

\usepackage{dcolumn}
\usepackage{bm}
\usepackage{color}

\makeatother

\usepackage{babel}

\begin{document}

\preprint{preprint(\today)}

\title{Charge order above room-temperature in a prototypical kagome superconductor La(Ru$_{1-x}$Fe$_{x}$)$_{3}$Si$_{2}$}

\author{I.~Plokhikh$^{\dag}$}
\email{igor.plokhikh@psi.ch}
\email{charles-hillis.mielke-iii@psi.ch} 
\affiliation{Laboratory for Multiscale Materials Experiments, Paul Scherrer Institut, CH-5232 Villigen PSI, Switzerland}

\author{C.~Mielke III$^{*}$}
\thanks{These authors contributed equally to the paper.}
\affiliation{Laboratory for Muon Spin Spectroscopy, Paul Scherrer Institute, CH-5232 Villigen PSI, Switzerland}
\affiliation{Physik-Institut, Universit\"{a}t Z\"{u}rich, Winterthurerstrasse 190, CH-8057 Z\"{u}rich, Switzerland}

\author{H.~Nakamura}
\affiliation{Institute for Solid State Physics (ISSP), University of Tokyo, Kashiwa, Chiba 277-8581, Japan}

\author{V.~Petricek}
\affiliation{Institute of Physics CAS, Na Slovance 1999/2, Praha, Czech Republic}

\author{Y.~Qin}
\affiliation{Wuhan National High Magnetic Field Center and School of Physics, Huazhong University of Science and Technology, Wuhan 430074, China}

\author{V.~Sazgari}
\affiliation{Laboratory for Muon Spin Spectroscopy, Paul Scherrer Institute, CH-5232 Villigen PSI, Switzerland}

\author{J.~K\"{u}spert}
\affiliation{Physik-Institut, Universit\"{a}t Z\"{u}rich, Winterthurerstrasse 190, CH-8057 Z\"{u}rich, Switzerland}

\author{I.~Bia\l{}o}
\affiliation{Physik-Institut, Universit\"{a}t Z\"{u}rich, Winterthurerstrasse 190, CH-8057 Z\"{u}rich, Switzerland}
\affiliation{AGH University of Science and Technology, Faculty of Physics and Applied Computer Science, 30-059 Krak\'{o}w, Poland}

\author{S.~Shin}
\affiliation{Laboratory for Multiscale Materials Experiments, Paul Scherrer Institut, 5232, Villigen PSI, Switzerland}

\author{O.~Ivashko}
\affiliation{Deutsches Elektronen-Synchrotron DESY, Notkestraße 85, 22607 Hamburg, Germany}

\author{M.v.~Zimmermann}
\affiliation{Deutsches Elektronen-Synchrotron DESY, Notkestraße 85, 22607 Hamburg, Germany}

\author{M.~Medarde}
\affiliation{Laboratory for Multiscale Materials Experiments, Paul Scherrer Institut, 5232, Villigen PSI, Switzerland}

\author{A.~Amato}
\affiliation{Laboratory for Muon Spin Spectroscopy, Paul Scherrer Institute, CH-5232 Villigen PSI, Switzerland}

\author{R.~Khasanov}
\affiliation{Laboratory for Muon Spin Spectroscopy, Paul Scherrer Institute, CH-5232 Villigen PSI, Switzerland}

\author{H.~Luetkens}
\affiliation{Laboratory for Muon Spin Spectroscopy, Paul Scherrer Institute, CH-5232 Villigen PSI, Switzerland}

\author{M.H. Fischer}
\affiliation{Physik-Institut, Universit\"{a}t Z\"{u}rich, Winterthurerstrasse 190, CH-8057 Z\"{u}rich, Switzerland}

\author{M.Z. Hasan}
\affiliation{Laboratory for Topological Quantum Matter and Advanced Spectroscopy (B7), Department of Physics,
Princeton University, Princeton, New Jersey 08544, USA}

\author{J.-X.~Yin}
\affiliation{Laboratory for Topological Quantum Matter and Advanced Spectroscopy (B7), Department of Physics,
Princeton University, Princeton, New Jersey 08544, USA}

\author{T. Neupert}
\affiliation{Physik-Institut, Universit\"{a}t Z\"{u}rich, Winterthurerstrasse 190, CH-8057 Z\"{u}rich, Switzerland}

\author{J.~Chang}
\affiliation{Physik-Institut, Universit\"{a}t Z\"{u}rich, Winterthurerstrasse 190, CH-8057 Z\"{u}rich, Switzerland}

\author{G.~Xu}
\affiliation{Wuhan National High Magnetic Field Center and School of Physics, Huazhong University of Science and Technology, Wuhan 430074, China}

\author{S.~Nakatsuji}
\affiliation{Institute for Solid State Physics (ISSP), University of Tokyo, Kashiwa, Chiba 277-8581, Japan}

\author{E.~Pomjakushina}
\affiliation{Laboratory for Multiscale Materials Experiments, Paul Scherrer Institut, CH-5232 Villigen PSI, Switzerland}

\author{D.J.~Gawryluk}
\email{dariusz.gawryluk@psi.ch}
\affiliation{Laboratory for Multiscale Materials Experiments, Paul Scherrer Institut, CH-5232 Villigen PSI, Switzerland}

\author{Z.~Guguchia}
\email{zurab.guguchia@psi.ch} 
\affiliation{Laboratory for Muon Spin Spectroscopy, Paul Scherrer Institute, CH-5232 Villigen PSI, Switzerland}

\maketitle

\textbf{The kagome lattice \cite{Syozi} is an intriguing and rich platform \cite{JiaxinNature,LYe,Mazin,GuguchiaCSS} for discovering, tuning and understanding the diverse phases of quantum matter, which is a necessary premise for utilizing quantum materials in all areas of modern and future electronics in a controlled and optimal way. The system LaRu$_{3}$Si$_{2}$ \cite{Barz,Vandenberg} was shown to exhibit typical kagome band structure features near the Fermi energy formed by the Ru-$dz^{2}$ orbitals \cite{GuguchiaPRM} and the highest superconducting transition temperature $T_{\rm c}$~${\simeq}$~7~K among the kagome-lattice materials. However, the effect of electronic correlations on the normal state properties remains elusive. Here, we report the discovery of charge order in La(Ru$_{1-x}$Fe$_{x}$)$_{3}$Si$_{2}$ ($x$~=~0,~0.01,~0.05) beyond room-temperature. Namely, single crystal X-ray diffraction reveals charge order with a propagation vector of ($\frac{1}{4}$,~0,~0) below $T_{\rm CO-I}$~${\simeq}$~400~K in all three compounds. At lower temperatures, we see the appearance of a second set of charge order peaks with a propagation vector of ($\frac{1}{6}$,~0,~0). The introduction of Fe, which is known to quickly suppress superconductivity \cite{GuguchiaPRM}, does not drastically alter the onset temperature for charge order. Instead, it broadens the scattered intensity such that diffuse scattering appears at the same onset temperature, however does not coalesce into sharp Bragg diffraction peaks until much lower in temperature. Our results present the first example of a charge ordered state at or above room temperature in the correlated kagome lattice with bulk superconductivity.}

There are three material classes of kagome lattice-systems which were recently shown to exhibit charge order: the $A$V$_{3}$Sb$_{5}$ ($A$~=~K, Rb, Cs) \cite{BOrtiz2,BOrtiz3,QYin} family of materials, ScV$_{6}$Sn$_{6}$ \cite{Arachchige,GuguchiaSc166,Bernevig,YongMing} and FeGe \cite{JiaxinPRL,XTeng}. $A$V$_{3}$Sb$_{5}$ exhibits two correlated orders: charge order below $T_{\rm co}$~${\simeq}$~80-110~K and a superconducting instability below $T_{\rm c}$~${\simeq}$~1-3~K \cite{YJiang,NShumiya,Wang2021}. The system ScV$_{6}$Sn$_{6}$ has a similar vanadium structural motif as $A$V$_{3}$Sb$_{5}$ and exhibits charge order below $T_{\rm co}$~${\simeq}$~90~K, however it is not superconducting down to the lowest measured temperature. FeGe is a correlated magnetic kagome system and exhibits an A-type AFM order below 400~K and charge order below 100~K. The unique features of $A$V$_{3}$Sb$_{5}$ and ScV$_{6}$Sn$_{6}$ are the emergence of possibly time-reversal symmetry (TRS) breaking chiral charge orders with both magnetic \cite{GuguchiaMielke,GuguchiaRVS,GuguchiaNPJ,KhasanovCVS,LiYu,YHu,YXu,GuoMoll,QWu,HuY,LiH,Saykin,YXing,TNeupert,JiaxinNature,GuguchiaSc166} and electronic anomalies \cite{SYang,FYu}. Theoretically, these features could be explained by a complex order parameter realizing a higher angular momentum state, dubbed unconventional \cite{MDenner,MHChristensen,MHChristensen2022,ERitz,Tazai2022,YHu,Balents,Nandkishore,Qimiao,DSong,Grandi,Tazai}, in analogy to superconducting orders. 

\begin{figure*}[t!]
\centering
\includegraphics[width=1.03\linewidth]{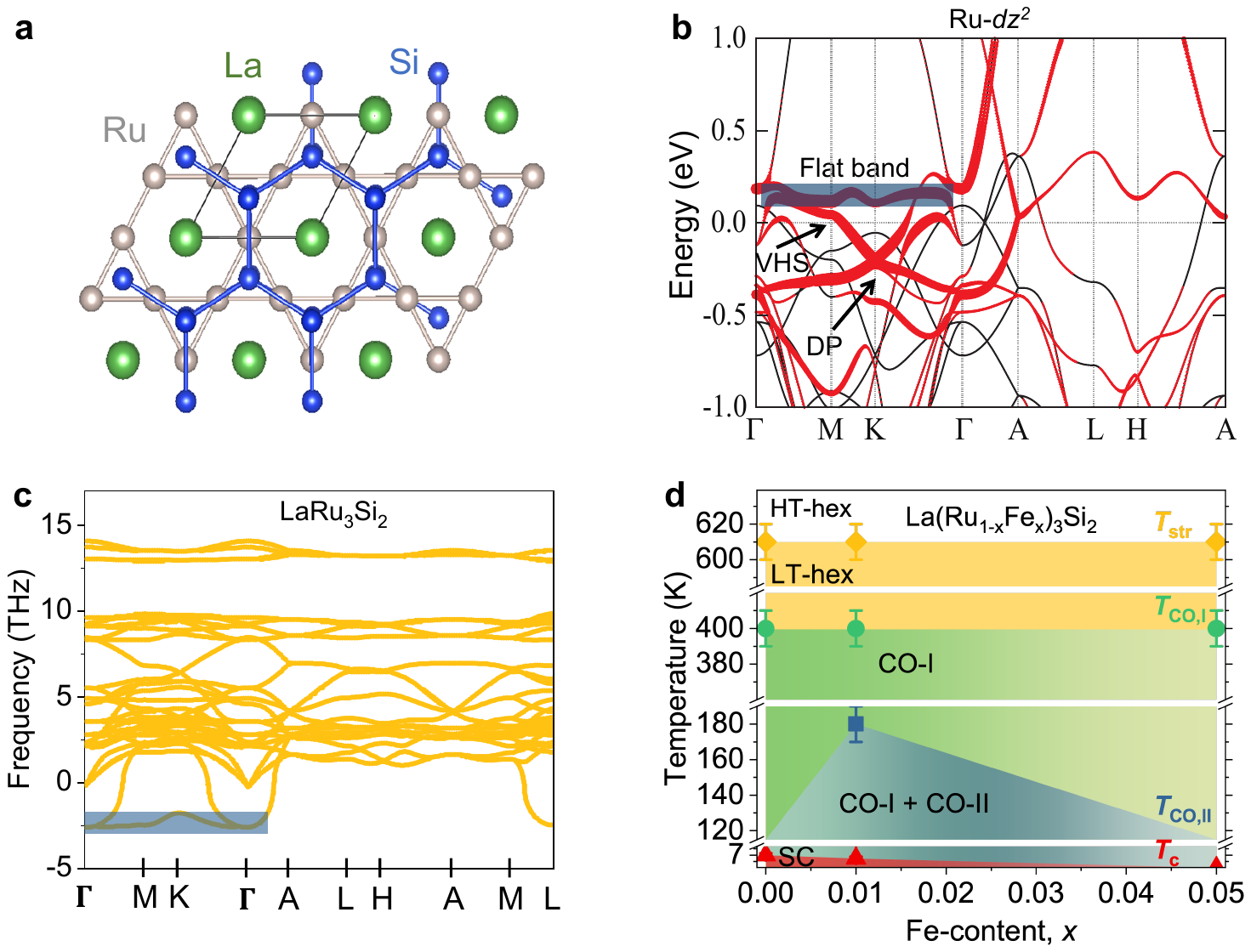}
\vspace{-1.2cm}
\caption{\textbf{High-temperature crystal structure, band structure and phase diagram.} 
$\bf{a}$, Top view of the atomic structure of LaRu$_{3}$Si$_{2}$. The Ru atoms construct a kagome lattice (middle size circles), while the Si (small size circles) and La atoms (large size circles) form a honeycomb and triangular structure, respectively.  Structures were plotted using the VESTA visualization tool \cite{Vesta}. $\bf{b}$, The band structure (black) and orbital-projected band structure (red) for the Ru-$dz^{2}$ orbital without SOC along the high symmetry $k$-path, presented in conformal kagome Brillouin Zone (BZ). The width of the line indicates the weight of each component. The blue-colored region highlights the manifestation of the kagome flat band. Arrows mark Dirac point (DP) and van Hove singulrity point (VHS) formed by the Ru-$dz^{2}$ orbitals near the Fermi level. $\bf{c}$, The phonon dispersion in the bulk of LaRu$_{3}$Si$_{2}$, obtained from local density approximation. $\bf{d}$, The Fe-doping evolution of the superconducting transition temperature $T_{\rm c}$ (data are taken from our previous work Ref. \cite{GuguchiaLRS}), charge order transition temperatures $T_{\rm CO-I}$, $T_{\rm CO-II}$ and structural phase transition temperature $T_{\rm str}$.}
\label{fig1}
\end{figure*}

\begin{figure*}[t!]
\centering
\includegraphics[width=1.0\linewidth]{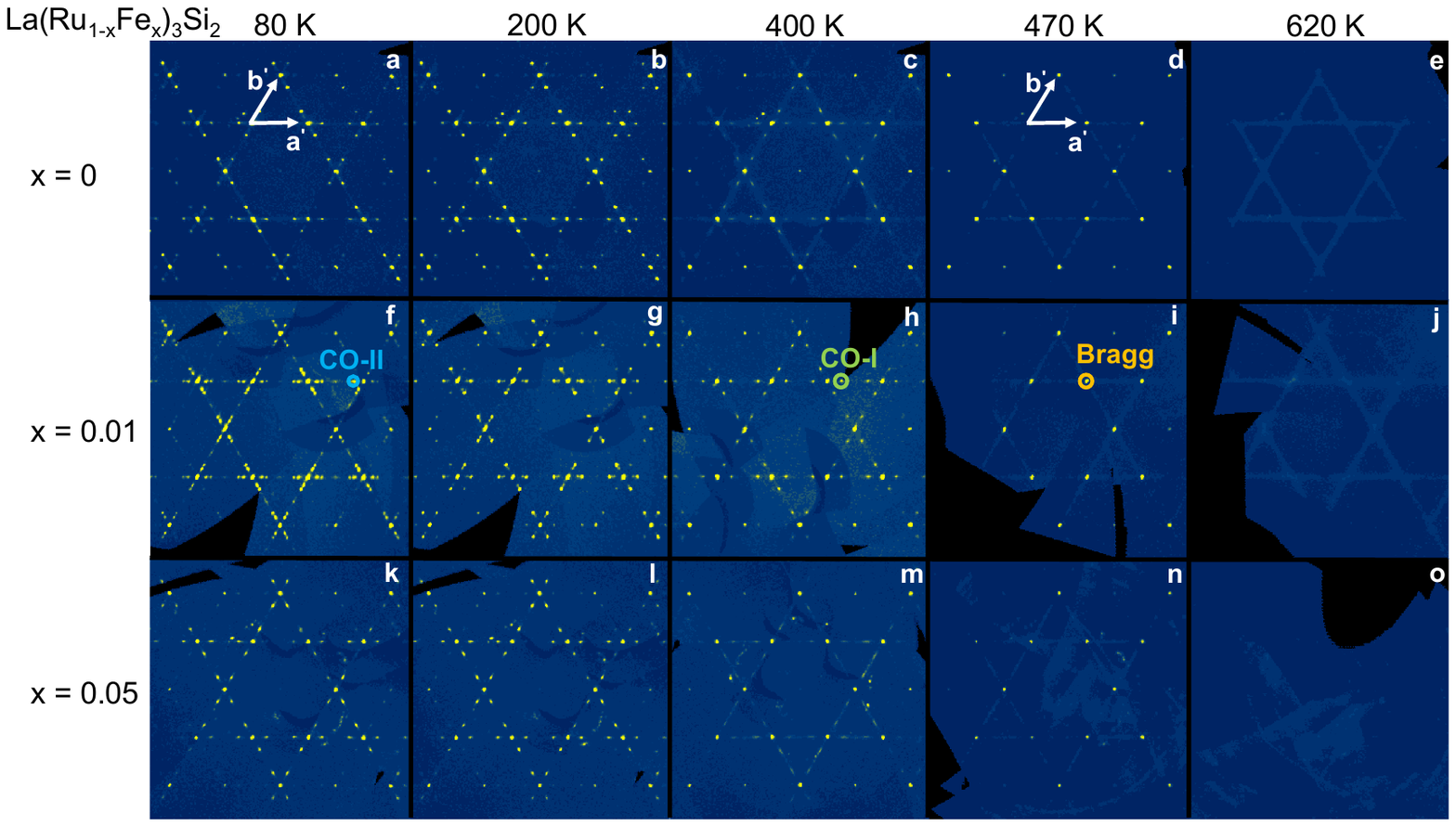}
\vspace{-1.5cm}
\caption{\textbf{Cascade of charge orders in La(Ru$_{1-x}$Fe$_{x}$)$_{3}$Si$_{2}$.} 
Reconstructed reciprocal space along the (0~0~1) direction at 3~r.l.u., performed at various temperatures for undoped $\bf{a-e,}$ as well as Fe-doped samples with $x$~=~0.01 $\bf{f-j,}$ and $x$~=~0.05 $\bf{k-o}$. Arrows indicate the reciprocal space vectors.}
\label{fig1}
\end{figure*}

A much less explored kagome lattice system is LaRu$_{3}$Si$_{2}$  \cite{GuguchiaPRM,Barz,Vandenberg,Kishimoto,LiWen,LiZeng}. The structure of LaRu$_{3}$Si$_{2}$ contains kagome layers of Ru sandwiched between layers of La and layers of Si having a honeycomb structure (see Fig. 1a)\cite{Barz,Vandenberg}. The superconductivity in LaRu$_{3}$Si$_{2}$ is fully gapped \cite{GuguchiaPRM} and has the highest SC transition temperature $T_{\rm c}$~${\simeq}$~7~K among the kagome-lattice materials. Using first principles calculations we previously found that the normal state band structure features a kagome flat band, Dirac point and van Hove point formed by the Ru-$dz^{2}$ orbitals near the Fermi level (see Fig. 1b). The electron-phonon coupling induced critical temperature $T_{\rm c}$, estimated from the phonon dispersion \cite{GuguchiaPRM}, was found to be much smaller than the experimental value. Thus, the enhancement of $T_{\rm c}$ in LaRu$_{3}$Si$_{2}$ was attributed to the presence of the flat band and the van Hove point relatively close to the Fermi level as well as to the high density of states from the narrow kagome bands \cite{GuguchiaPRM}. Regarding the normal state properties, there is no report on either magnetic order or charge order in this system. Anomalous properties \cite{LiZeng,Kishimoto} in the normal state were inferred from thermodynamic measurements in LaRu$_{3}$Si$_{2}$, such as the deviation of the normal state specific heat from the Debye model, non-mean field like suppression of superconductivity with magnetic field and non-linear field dependence of the induced quasiparticle density of states (DOS). Importantly, from first principles calculations we find a phonon dispersion with negative frequency modes \cite{GuguchiaPRM} in LaRu$_{3}$Si$_{2}$, as shown in Fig. 1c. The soft phonon band is relatively flat. So, the phonon calculations in LaRu$_{3}$Si$_{2}$ point to structural distortions due to negative frequency modes. Their soft dispersion hints at the possibility of charge order physics.  We note that imaginary (soft), almost-flat  phonon modes were also observed in kagome system ScV$_{6}$Sn$_{6}$ \cite{Bernevig}. It was also shown that imaginary phonon and the strong fluctuations of its flat band induce an unconventional charge order phase, which was observed experimentally \cite{Arachchige,GuguchiaSc166,YongMing}. 

In order to explore the possibility of the formation of charge order in the title compound, we investigated three microcrystalline samples of La(Ru$_{1-x}$Fe$_{x}$)$_{3}$Si$_{2}$ ($x$~=~0, 0.01, 0.05) as well as a large single crystal of LaRu$_{3}$Si$_{2}$ via X-ray diffraction (the experimental details of which may be found in the Supplementary Information). The experiments reveal charge order with a propagation vector of ($\frac{1}{4}$,~0,~0) in undoped LaRu$_{3}$Si$_{2}$ as well as Fe-doped La(Ru$_{1-x}$Fe$_{x}$)$_{3}$Si$_{2}$ samples with $x$~=~0.01 and $x$~=~0.05, setting in well above room temperature ($T_{\rm CO-I}$~${\simeq}$~400~K) (see Fig.~1d). The charge order transition is preceded by a structural phase transition with $T_{\rm str}$~${\simeq}$~600~K from the high-temperature hexagonal $P6/mmm$ phase (SG No.~191) to the low-temperature orthorhombic $Cccm$ phase (SG No.~66). Furthermore, a second set of charge order peaks with a propagation vector of ($\frac{1}{6}$,~0,~0) sets in below $T_{\rm CO-II}$~${\simeq}$~180~K in the $x$~=~0.01 sample and possibly in the $x$~=~0 sample below $T_{\rm CO-II}$~${\simeq}$~80~K, showing competition with the primary charge order.


\begin{figure*}[t!]
\centering
\includegraphics[width=1.0\linewidth]{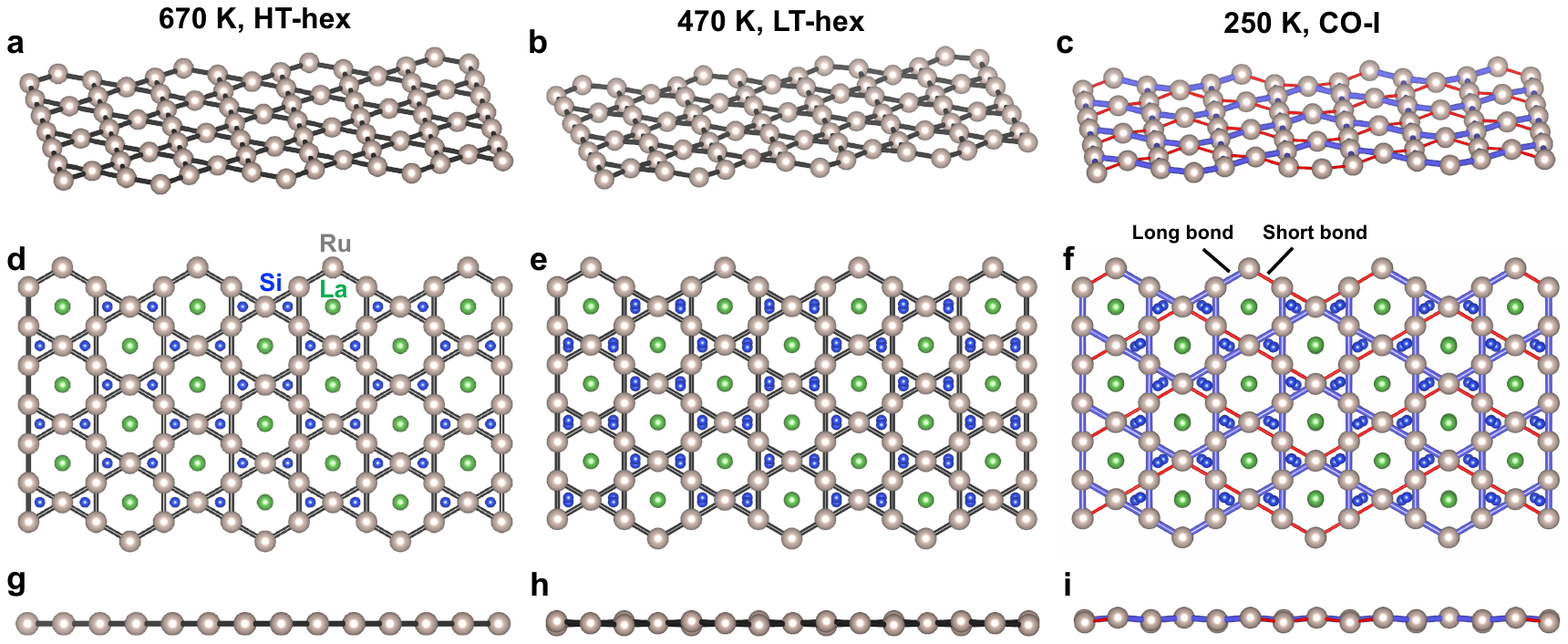}
\vspace{-4.3cm}
\caption{\textbf{Crystal structure.} 
General projection of the kagome net at 670~K $\bf{(a)}$, 470~K $\bf{(b)}$ and 250~K $\bf{(c)}$, projection perpendicular to the kagome net at 670~K $\bf{(d)}$, 470~K $\bf{(e)}$ and 250~K $\bf{(f)}$, projection along the kagome net at 670~K $\bf{(g)}$, 470~K $\bf{(h)}$ and 250~K $\bf{(i)}$.}
\label{fig1}
\end{figure*}

Previous reports for both pristine and doped LaRu$_3$Si$_2$ \cite{Barz,SLi2011,BLi2016, SLi2012} have some discrepancy regarding the room temperature structure, represented either by a small hexagonal cell ($P6/mmm$, $a$ ${\simeq}$~5.6~$\rm \AA$, $c$~${\simeq}$ 3.5~$\rm \AA$) (see supplementary Fig.~S5) or large hexagonal cell ($P6_{3}/m$, $a$ ${\simeq}$ 5.6~$\rm \AA$,~c~${\simeq}$ 7.1~$\rm \AA$) (see supplementary Fig.~S6). Above 600~K, we found, from our single crystal diffraction experiments, that the crystal structure can indeed be described using a small hexagonal cell ($P6/mmm$, $a\times a\times c$, \textbf{HT-hex}) (see supplementary Fig.~S5). The results of the refinement are summarized in the Supplementary Tables S1-S5 and provided as separate cif-files. Attempts to introduce Fe in the Ru position for the $x$~=~0.01 Fe-doped sample yields 1.6(1)\%, which agrees with the nominal value of 1\% Fe-doping. Another peculiar feature is presence of 5.6(2)\% vacancies in the Si position.

Although the high temperature model described above fits the diffraction data well, it features notably oblate thermal ellipsoids for Ru and Si atoms (Supplementary Table S5). This, together with pronounced diffuse scattering within the (0,~0,~$l$~=~odd) reciprocal space planes (see Figure 2$\bf{d}$,$\bf{e}$,$\bf{i}$,$\bf{j}$,$\bf{n}$), indicates the proximity to a phase transition. Indeed, upon cooling below 600~K, we observe evolving Bragg peaks corresponding to the doubling of the initial unit cell along the $c$-axis ($a\times a\times 2c$, \textbf{LT-hex}). The diffraction pattern follows hexagonal symmetry ($R_{int}$~=~1.24\% for $P6/mmm$ Laue symmetry) in agreement with the $P6_{3}/m$ space group  (SG No.~176), previously inferred from powder data \cite{Barz}. Attempts to solve the crystal structure in this space group or any other down to $P1$ are unsuccessful, which is likely an indication of twinning upon the phase transition. The structure can be described in the space group $Cccm$ (No.~66) as three interpenetrating orthorhombic twins ($a\times\sqrt3a\times2c$) propagated by the six-fold axis present in the parent high-temperature phase. The refined twin fractions are close to $\frac{1}{3}$, which effectively mimics the six-fold symmetry of the diffraction pattern.

\begin{figure}[t!]
\centering
\includegraphics[width=1.0\linewidth]{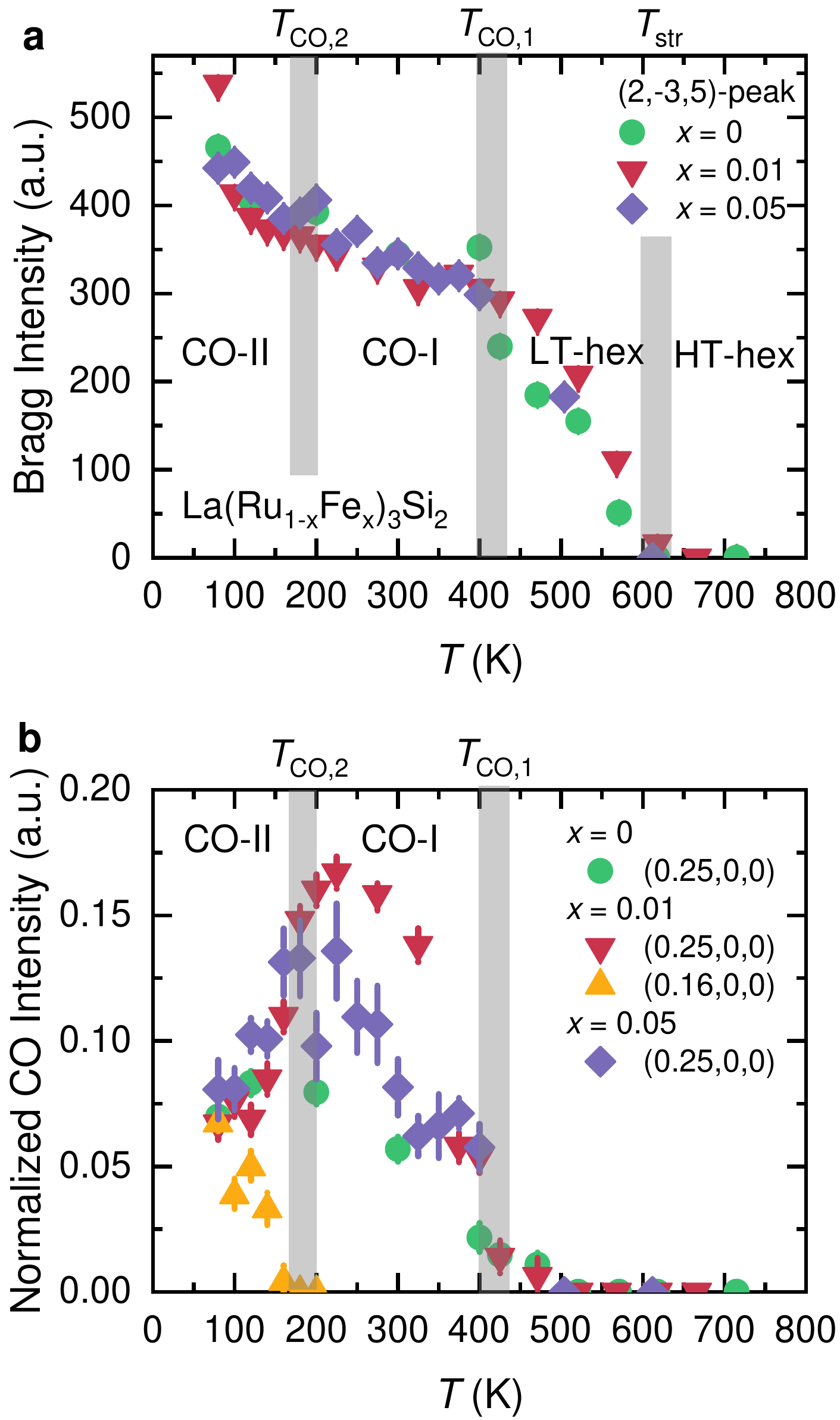}
\vspace{-0.5cm}
\caption{\textbf{Crystal structures and charge orders in La(Ru$_{1-x}$Fe$_{x}$)$_{3}$Si$_{2}$.} 
Temperature evolution of the Bragg intensity ($\bf{a}$) and charge order satellite peak intensities ($\bf{b}$ ).}
\label{fig1}
\end{figure}

Below $\sim$400~K, the diffuse scattering is clustered into another set of sharp diffraction spots. Besides the main reflections, there are satellites corresponding to $q_1$~=~($\frac{1}{4}$~0~0), $q_2$~=~(0~$\frac{1}{4}$~0) and $q_3$~=~($\frac{1}{4}~\frac{-1}{4}$~0) relative to the $a\times a\times 2c$ cell 
(see Fig. 2$\bf{e-c}$); the refined deviations from the rational fractions are within 3$\sigma$. The diffraction pattern thus represents a $4\times4\times1$ superstructure of the parent diffraction pattern existing above $\sim$400~K; all reflections can be indexed using a $4a\times4a\times2c$ supercell. Similar to the case of $\textbf{LT-hex}$, the structure can be solved in this supercell in an orthorhombic unit cell ($Cccm$, 40 symmetrically independent atoms) assuming twinning by the six-fold axis; the results are provided in a separate cif-file. The disadvantage of this description is that it also accounts for higher order satellite peaks besides $q_1$, $q_2$ and $q_3$,  including cross-satellites ($2q_1$, $q_{1}q_{2}$ etc.), which are practically unobservable. This hints at the possibility of describing the $\textbf{CO-I}$ phase with a smaller cell. Indeed, the initial supercell can be reduced to $a\times2\sqrt3a\times2c$ or $a\times\sqrt3a\times2c$ with $q$~=~(0~$\frac{1}{2}$~0) cell without losing observable reflections. Within this cell, the structure can be described as modulated in superspace group $Cmmm(0b0)s00$.
Superlattice peak intensity appears below $T_{\rm CO-I}$~${\simeq}$~400~K for undoped LaRu$_{3}$Si$_{2}$ Fig. 2$\bf{a-e}$ as well as Fe-doped La(Ru$_{1-x}$Fe$_{x}$)$_{3}$Si$_{2}$ samples with $x$~=~0.01 and $x$~=~0.05 as shown in Figures 2$\bf{a-e}$, 2$\bf{f-j}$, and 2$\bf{k-o}$, respectively.
The charge order features for $x$~=~0.05 sample looks broader, as compared to $x$~=~0 and 0.01 samples and do not coalesce into sharp diffraction peaks right below ${\simeq}$~400~K. This indicates that the effect of Fe-doping is to introduce disorder into the charge order. Interestingly, another set of reflections evolves at positions corresponding to $q_{1}'$~=~($\frac{1}{6}~0~0$), $q_{2}'$~=~(0~$\frac{1}{6}~0$) and $q_{3}'$~=~($\frac{1}{6}~\frac{-1}{6}~0$), which is most pronounced for the $x$~=~0.01 sample with $T_{\rm CO-II}$~${\simeq}$~180~K. Since the relation between these two sets of ordering vectors is unclear (multi-phase vs. multi-$q$ ordering) we refrain from speculating on the possible models of superstructure ordering.

Comparison of models for $\textbf{HT-hex}$ (at 670~K), $\textbf{LT-hex}$ (at 470~K) and $\textbf{CO-I}$ (at 250~K) phases is provided in Figure 3$\bf{a-i}$. The $\textbf{HT-hex}$ phase features an undistorted planar kagome net comprised by Ru atoms alternating with Si/La planes. The distances within the Ru-kagome net are 2.844(4)$\rm~\AA$. Alternatively, this structure can be viewed as packing of La polyhedra comprised by 6 Si atoms (d(La-Si) = 3.284(5)$\rm \AA$) and 12 Ru atoms (d(La-Ru)~=~3.357(3)$\rm~\AA$) reinforced by short Si-Ru contacts (d(Si-Ru)~=~2.424(6)$\rm~\AA$). The transition to the $\textbf{LT-hex}$ phase is mainly due to in-plane displacement of Si atoms and out-of-plane displacements of 2/3 of Ru atoms. In particular, this leads to corrugations of the ideal kagome net with shortening and elongation of Ru-Ru distances (4$\times$2.8431(7)$\rm~\AA$~+~2$\times$2.853(1)$\rm~\AA$), but still retaining a nearly ideal hexagonal angle. The environment of La is also distorted. Upon further cooling and transition into the $\textbf{CO-I}$ phase, there occurs further disproportionation of bonds; the Ru-Ru distances are in the range 2.832(1)~-~2.868(1)$\rm~\AA$. In Figure 3$\bf{c}$, $\bf{f}$, $\bf{i}$, the distances below 2.845$\rm~\AA$ are marked as short and the distances above 2.845$\rm~\AA$ are marked as long.

Figures 4$\bf{a}$ and $\bf{b}$ outlines four distinct temperature regions for La(Ru$_{1-x}$Fe$_{x}$)$_{3}$Si$_{2}$ by following selected main and supposedly superstructure reflections. Namely, Figs. 4$\bf{a}$ and $\bf{b}$ show the temperature dependences of the intensities of the (2~-3~5) (hexagonal setting) Bragg peak and the charge order peaks intensities, respectively, for $x$~=~0, 0.01 and 0.05 samples. By tracking the relative intensity of $\textbf{CO-I}$ and $\textbf{CO-II}$ peaks for the different doped samples, we find that in the $x$~=~0.01 case, there is a clear decrease in the scattered intensity from the $\textbf{CO-I}$ phase at the same temperature where there appears the scattered intensity from the $\textbf{CO-II}$ phase. The intensity of the peaks from the $\textbf{CO-II}$ phase increases while the intensity of the peaks from the $\textbf{CO-I}$ monotonously decrease, until they achieve equivalent intensities at 80~K. This implies that the two charge ordered phases compete. We note that the onset of this additional modulated structure $\textbf{CO-II}$ appears to be enhanced in the $x$~=~0.01 Fe-doped sample, occurring at 180~K, while weak peaks become visible below 120-100~K in the pristine sample. The intensity of the $\textbf{CO-I}$ phase seems to reach a peak at 120~K and decrease from 120~K to 80~K in the undoped material, suggesting that the $\textbf{CO-II}$ phase may onset at lower temperatures. No signatures of coherent or diffuse scattering at $\frac{1}{6}$ periodicity is observed in the $x$~=~0.05 Fe-doped sample, which is most likely due to disorder-induced broadening of charge order peaks. The intensities of the (2~-3~5) (hexagonal setting) Bragg peak for $x$~=~0, 0.01 and 0.05 samples (see Figure 4$\bf{a}$) show a clear slope change around $T_{\rm CO-I}$~${\simeq}$~400~K and an additional slope change with an increase in (2~-3~5) intensity below $T_{\rm CO-II}$~${\simeq}$~180~K for all three samples. Increase of Bragg peak intensity across $T_{\rm CO-II}$ is most pronounced in the $x$~=~0.01 sample in which a clear indication of $\frac{1}{6}$ charge order is observed. Additional low temperature measurements are crucial to gain insight into the low temperature charge order phase CO-II.

Experimental realisation of charge order in the kagome lattice has long been awaited until recently when charge order was found in the family of kagome metals $A$V$_{3}$Sb$_{5}$ ($A$~=~K, Rb, Cs) \cite{BOrtiz2,BOrtiz3,QYin,YJiang,NShumiya,Wang2021}, ScV$_{6}$Sn$_{6}$ \cite{Arachchige,GuguchiaSc166} and magnetic FeGe \cite{JiaxinPRL,XTeng}. However, the charge ordering temperature in all these materials are limited to ${\simeq}$~100~K, i.e., well below room temperature. Also, among the above-mentioned systems it is only the $A$V$_{3}$Sb$_{5}$ family of materials which shows superconductivity. In these systems, charge order strongly competes with superconductivity and has a negative effect on both the superconducting critical temperature and the superfluid density. Therefore, the maximum SC transition temperature, in the presence of charge order, was reported to be ${\simeq}$~3~K. Here, we report on the discovery of the cascade of charge orders in La(Ru$_{1-x}$Fe$_{x}$)$_{3}$Si$_{2}$ ($x$~=~0-0.05): the one with a propagation vector of ($\frac{1}{4}$,~0,~0), setting in well above room temperature ($T_{\rm CO-I}$~${\simeq}$~400~K) and a second one with the propagation vector of ($\frac{1}{6}$,~0,~0), setting at lower temperatures, which competes with the primary charge order. Moreover, the system LaRu$_{3}$Si$_{2}$ exhibits the highest superconducting critical temperature $T_{\rm c}$ ${\simeq}$~7~K and the highest superfluid density \cite{GuguchiaPRM} among the known bulk kagome-lattice superconductors. This implies that in this system both charge order and superconductivity are somewhat optimised.  Fe-doping does introduce significant disorder to the charge-ordered state and leads to the full suppression of superconductivity with $x_{\rm Fe, \textit{cr}}$~=~0.05 \cite{GuguchiaLRS}. Our results classify the bulk superconductor LaRu$_{3}$Si$_{2}$ as the first kagome-lattice system with charge order above room temperature. 

Room-temperature charge order will be useful in charge order-based electronic devices \cite{GLiu,Balandin}. Switching between various material phases near room temperature is a promising step toward electronic and optoelectronic technologies. For instance, the abrupt change in resistance during charge order phase transition can be used to achieve circuits and logic gates without the need for field-effect transistors. There is an example of an oscillator based on an integrated TaS$_{2}$-boron nitride-graphene device, which utilizes the charge order phase transition \cite{GLiu}. Previous works have also discussed the implementation of charge order transition based devices \cite{Balandin} for information processing and radiation-hard applications. In theory, such devices can be fast, low-power, and immune to proton and X-ray radiation.

Having established the existence of high-temperature charge order in the kagome material LaRu$_{3}$Si$_{2}$, our study opens up exciting avenues for future research. In the kagome compounds ScV$_{6}$Sn$_{6}$ \cite{Arachchige,GuguchiaSc166,Bernevig,YongMing} and $A$V$_{3}$Sb$_{5}$ ($A$~=~K, Rb, Cs)  \cite{TNeupert,YJiang,GuguchiaMielke,GuguchiaRVS,MDenner,MHChristensen2022,Wang2013,Brien2010,Kiesel2013} the interplay of electronic and phononic contributions in the formation of charge order have been at the heart of the scientific discourse. It seems clear that both the low-energy fermiology as well as the phononic instabilities have to be considered to explain the charge order with its unconventional properties, such as chiral charge order that can be sensitively manipulated by strain and magnetic fields. While charge order in this context means a translation symmetry breaking instability ($\boldsymbol{Q}\neq0$ order), there exists the equally relevant possibility of an intra-unit cell ($\boldsymbol{Q}=0$) charge order in which low energy electronic states involved. The loop current phases of the Haldane honeycomb model \cite{Haldane} or the Varma model \cite{Varma} for the cuprate superconductors are prominent examples. Following this line of thought blurs the boundaries between structural transitions and charge orders. It is thus imparative to study the electronic response and electronic structure of LaRu$_{3}$Si$_{2}$ in more detail to unveil its connections with the changes of structural motifs found in this work.

\textbf{Acknowledgment}
Z.G. acknowledges support from the Swiss National Science Foundation (SNSF) through SNSF Starting Grant (No. TMSGI2${\_}$211750). We acknowledge DESY (Hamburg, Germany), a member of the Helmholtz Association HGF, for the provision of experimental facilities. Parts of this research were carried out at P21.1 beamline. I.P. acknowledges support from Paul Scherrer Institute research grant No. 2021${\_}$01346 and S.S. acknowledges support from the SNSF through grant  
No. 200021${\_}$188706. I.P. acknowledges Dr. Vladimir Pomjakushin for valuable discussion.\\ 

\textbf{Author Contributions}
Z.G. conceived and supervised the project. Growth of La(Ru$_{1-x}$Fe$_{x}$)$_{3}$Si$_{2}$ samples: C.M.III., and D.J.G.. Growth of undoped sample LaRu$_{3}$Si$_{2}$: H.N., and S.N.. X-ray diffraction experiments at DESY, analysis and corresponding discussions: C.M.III, J.K., I.B., O.I., M.v.Z., J.C., and Z.G.. Laboratory single crystal X-ray diffraction experiments and corresponding discussions: I.P., C.M.III, V.P., E.P., S.S., M.M., D.J.G., and Z.G.. DFT calculations and corresponding discussions: Y.Q., G.X., T.N., M.H.F., J.-X.Y., M.Z.H., and Z.G.. X-ray data refinement: I.P. and V.P.. Figure development and writing of the paper: Z.G., I.P., C.M.III., and D.J.G. with contributions from all authors. All authors discussed the results, interpretation and conclusion.\\

\textbf{Competing interests:} All authors declare that they have no competing interests.\\

\newpage

\section{METHODS}

\textbf{Sample preparation}:
The samples of La(Ru$_{1-x}$Fe$_{x}$)$_{3}$Si$_{2}$ were synthesized from high-purity Si lump (purity 99.999+\%, Alpha aeser), low-oxygen vacuum remelted Fe lump (purity 99.99\%, Alpha aeser), Ru pellets (purity 99.95\%, Alpha aeser), and La ingot (purity 99.9\%) via arc melting using Zr pellets as an oxygen getter. The melted buttons were flipped three times to ensure melt homogenization. In order to suppress the second phase LaRu$_2$Si$_2$, an additional 15\% Ru was added to each melt (as mentioned previously \cite{SLi2011, SLi2012, BLi2016}) in order to attain nominal compositions as La(Ru$_{1-x}$Fe$_{x}$)$_{3.45}$Si$_{2}$. Single crystals of La(Ru$_{1-x}$Fe$_{x}$)$_{3}$Si$_{2}$ were extracted from arc-melted buttons.\\ 


\textbf{Single crystal diffraction measurements}: Single crystalline samples were selected from the large undoped crystal and from crushed chunks of the arc-melted samples washed in dilute HCl for 5 minutes before being flushed in deionized H$_2$O and ethanol followed by acetone to dry them. Following this, we selected single crystalline samples which had faceted sides and a typical size of 30~to~100~$\mu$m. These were then mounted on a small-diameter (10~$\mu$m) MiTeGen loop or a quartz capillary and centered within the X-ray beam. Diffraction measurements were carried out using the laboratory $STOE$ STADIVARI single crystal diffractometer at the Paul Scherrer Institute. Measurements were carried out in the temperature range between 80~K and 700~K, using the combination of the CryoStream (constant N$_2$ gas flow) and HeatStream (constant Ar flow of 0.8~L/min). The micro-focused source was a Mo $K_\alpha$ X-ray, with a wavelength of 0.71073~$\rm \AA$. The detector was a Dectris EIGER 1M 2R. The exposure time was set to 10~seconds per frame, and the coverage was calculated through the program X-Area [X-Area package, Version 2.1, STOE and Cie GmbH, Darmstadt, Germany, 2022]. Data reduction and correction was done with X-Area package. The peaks were identified by the X-Area software (typically taking a minimum Intensity/$\sigma~\geq$~25) and then the lattice parameters were refined. From this, we were able to integrate the intensities and extract individual Bragg peak and CO peak intensities for each $q$-vector (modulation). These intensities were corrected by beam-out and frame scaling, and then furthermore by spherical correction (where the absorption $\mu$ and sample radius were inputs). From this, we were able to identify the strongest CO satellite intensities and follow them with temperature. We also used the X-Area software to perform reconstructions in reciprocal space of the collected data (see Figure 2) in order to visually identify changes in the charge order periodicity and structure. We used the software JANA2020 for data refinement and $\mu$ calculation \cite{JANA2020}. The crystal of LaRu$_{3}$Si$_{2}$ was also measured at room temperature in the hard X-Ray beamline (P21.1, PETRA III) at the DESY facility and charge order was identified (Figure 8).

\begin{table*}
\caption{Details of X-ray diffraction experiments for $x=$~0.01 Fe-doped La(Ru$_{1-x}$Fe$_{x}$)$_3$Si$_2$.}

\begin{tabular}{| c |c |c |c|}
 \hline
  & \textbf{HT-hex} & \textbf{LT-hex} & \textbf{Co-I}\\
 \hline
  Radiation & \multicolumn{3}{|c|}{Mo$K_{\alpha}$, $\lambda$~=~0.71073$\rm \AA$}\\
 \hline
 Temperature & 670~K & 470~K & 250~K\\
 \hline
 Space Group & $P6/mmm$ (No. 191) & $Cccm$ (No. 66) & $Cmmm(0b0)s00$ \\
 \hline
 $a$ in $\AA$ & 5.688(3) & 5.6794(1) & 5.6738(9)\\
 $b$ in $\AA$ & =$a$ & 9.8370(1) & 9.827(2)\\
 $c$ in $\AA$ & 3.567(3) & 7.109(2) & 7.1124(9)\\
  $k$ &  &  & (0 $\frac{1}{2}$ 0)\\
 $V$ in $\AA^3$ & 99.9(1) & 397.2(2) & 396.6(2)\\
 \hline
 $D_{calc}$ in g/cm$^3$ & 8.1933 & 8.1949 & 8.2116\\
 \hline
 $\mu$ in cm$^{-1}$ & 21.911 & 22.051 & 22.038\\
 \hline
  $\theta$ range in degree & 4.14 - 29.51 & 4.14 - 29.96 & 2.86 - 33.16\\
 \hline
  Reflections measured & 467 & 1310 & 42519 \\
  Averaged reflections all & 78 & 509 &  2872\\
  Averaged reflections I/3$\sigma$ & 76 & 453 & 1873 \\
 \hline
 $hklm$ range & -7 \textless $h$ \textless 7  & -7 \textless $h$ \textless 7  & -33 \textless $h$ \textless34\\
   & -4 \textless $k$ \textless 7  & -9 \textless $k$ \textless13  & -60 \textless $k$\textless 59\\
   & -4 \textless $l$ \textless4  & -9 \textless $l$ \textless13  & -10 \textless $l$ \textless10\\
      &   &  & -1 \textless $m$ \textless1\\
 \hline
$\Delta \rho_{max}$/$\Delta \rho_{min}$ in e$^-$ & 0.51/-0.70 & 1.33/-1.77 & 1.20/-1.01\\
  \hline
 $\rm R_{int}$, $\textless \sigma / I \textgreater$ in \% & 0.72/0.17 & 1.03/0.52 & 1.69/0.28 \\
  \hline
   $\chi^{2}$($I\textgreater3\sigma$), $\chi^2$(all) & 1.51/1.47 & 1.07/1.06 & 3.64/2.96\\
  \hline
   $\rm R_{F}$($I\textgreater3\sigma$), $\rm R_{F}$(all) & 1.46/1.52 & 1.29/2.01 & 2.85/4.24 \\
  \hline
    w$\rm R_{F}$($I\textgreater3\sigma$), w$\rm R_{F}$(all)  & 2.96/3.08 & 2.02/2.18 & 5.16/5.22 \\
  \hline
\end{tabular}
\end{table*}

\begin{table}
\caption{Twin Matrices at 470 K from refinements in $Cccm$ space group.}
\begin{tabular}{ c c c }
 \hline
 \multicolumn{3}{|c|}{Twin 1, 30.7(5)\%} \\
 \hline
  1 & 0 & 0 \\
  0 & 1 & 0 \\
  0 & 0 & 1 \\
  \hline
\end{tabular}

\begin{tabular}{ c c c }
 \hline
 \multicolumn{3}{|c|}{Twin 2, 37.2(3)\%} \\
 \hline
  $\frac{1}{2}$ & $\frac{-3}{2}$ & 0 \\
  $\frac{1}{2}$ & $\frac{1}{2}$ & 0 \\
  0 & 0 & 1 \\
  \hline
\end{tabular}

\begin{tabular}{ c c c }
 \hline
 \multicolumn{3}{|c|}{Twin 3, 32.1(3)\%} \\
 \hline
  $\frac{-1}{2}$ & $\frac{-3}{2}$ & 0 \\
  $\frac{1}{2}$ & $\frac{-1}{2}$ & 0 \\
  0 & 0 & 1 \\
  \hline
\end{tabular}
\end{table}

\begin{table}
\caption{Twin matrices at 250 K in the $Cmmm(0b0)s00$ setting.}

\begin{tabular}{ c c c }
 \hline
 \multicolumn{3}{|c|}{Twin 1, 27.4(2)\%} \\
 \hline
  1 & 0 & 0 \\
  0 & 1 & 0 \\
  0 & 0 & 1 \\
  \hline
\end{tabular}

\begin{tabular}{ c c c }
 \hline
 \multicolumn{3}{|c|}{Twin 2, 28.2(1)\%} \\
 \hline
  $\frac{1}{2}$ & $\frac{3}{2}$ & 0 \\
  $\frac{-1}{2}$ & $\frac{1}{2}$ & 0 \\
  0 & 0 & 1 \\
  \hline
\end{tabular}

\begin{tabular}{ c c c }
 \hline
 \multicolumn{3}{|c|}{Twin 3, 44.3(1)\%} \\
 \hline
  $\frac{1}{2}$ & $\frac{3}{2}$ & 0 \\
  $\frac{1}{2}$ & $\frac{-1}{2}$ & 0 \\
  0 & 0 & 1 \\
  \hline
\end{tabular}
\end{table}

\begin{table*}
\caption{Atomic coordinates and displacement parameters for $x=$~0.01 Fe-doped La(Ru$_{1-x}$Fe$_{x}$)$_3$Si$_2$.}
\begin{tabular}{ ||c |c| c| c| c| c| c||}
 \hline
  Atom & Wyckhoff & x/a & y/b & z/c & Occupancy & $U_{iso}$\\
\hline
 \hline
 \multicolumn{7}{|c|}{\textbf{HT-hex}, $P6/mmm$, $a\times a\times c$} \\
 \hline
  \hline
 La1 & 1$b$ & 0 & 0 & 1/2 & 1 & 0.0217(4)\\
  \hline
 Ru1/Fe1 & 3$f$ & 1/2 & 0 & 0 & 0.984(1)/0.016(1) & 0.0242(4)\\
  \hline
 Si1 & 2$d$ & 2/3 & 1/3 & 1/2 & 0.944(2) & 0.030(1)\\
 \hline
  \hline
 \multicolumn{7}{|c|}{\textbf{LT-hex}, $Cccm$, $a\times \sqrt3a\times 2c$} \\
 \hline
  \hline
 La1 & 4$d$ & 1/2 & 0 & 0 & 1 & 0.0183(4)\\
  \hline
 Ru1/Fe1 & 4$a$ & 1/2 & 1/2 & 1/4 & 0.862(7)/0.138(7) & 0.0187(4)\\
  \hline
 Ru2 & 8$k$ & 1/4 & 3/4 & 0.23058(7) & 1 & 0.0175(2)\\ 
  \hline
 Si1 & 8$l$ & 0.5291(3) & 0.6680(4) & 0 & 0.963(7) & 0.0205(8)\\
 \hline
  \hline
 \multicolumn{7}{|c|}{\textbf{CO-I}, $a\times \sqrt3a\times 2c$, $Cmmm(0b0)s00~q~=~(0~\frac{1}{2}~0)$} \\
 \hline
  \hline
 La1 & - & 1/2 & 1/2 & 0 & 1 & 0.0066(2)\\
 $sin$ &  & 0.00803(9) & 0 & 0 &  & \\
  $cos$ &  & 0 & 0 & 0 &  & \\
  \hline
 La2 & - & 1/2 & 1/2 & 1/2 & 1 & 0.0065(2)\\
  $sin$ &  & -0.00914(9) & 0 & 0 &  & \\
  $cos$ &  & 0 & 0 & 0 &  & \\
  \hline
 Ru1 & - & 0 & 1/2 & 0.2252(1) & 1 & 0.0060(2)\\
  $sin$ &  & -0.00133(9) & 0 & 0 &  & \\
  $cos$ &  & 0 & 0 & 0 &  & \\
  \hline
 Ru2/Fe2 & - & 1/4 & 3/4 & 0.26168(5) & 0.930(4)/0.070(4) & 0.0070(1)\\
  $sin$ &  & -0.0014(1) & 0.00143(3) & 0 &  & \\
  $cos$ &  & 0 & 0 & 0 &  & \\
  \hline
 Si1 & - & 0 & 0.6839(2) & 0 & 1 & 0.0089(5)\\
  $sin$ &  & 0.0081(3) & 0 & 0 &  & \\
  $cos$ &  & -0.0325(4) & 0 & 0 &  & \\
  \hline
 Si1 & - & 0 & 0.6473(2) & 1/2 & 0.953(6) & 0.0075(5)\\
  $sin$ &  & -0.0078(3) & 0 & 0 &  & \\
  $cos$ &  & 0.0194(3) & 0 & 0 &  & \\
 \hline
\end{tabular}
\end{table*}

\begin{table*}
\caption{Anisotropic atomic displacement parameters for $x=$~0.01 Fe-doped La(Ru$_{1-x}$Fe$_{x}$)$_3$Si$_2$.}
\begin{tabular}{ ||c |c| c| c| c| c| c||}
 \hline
  Atom & $U_{11}$ & $U_{22}$ & $U_{33}$ & $U_{12}$ & $U_{13}$ & $U_{23}$\\
\hline
 \hline
 \multicolumn{7}{|c|}{\textbf{HT-hex}, $P6/mmm$, $a\times a\times c$} \\
 \hline
  \hline
 La1 & 0.0217(5) & $=U_{11}$ & 0.0215(6) & 0.0109(3) & 0 & 0\\
  \hline
 Ru1/Fe1 & 0.0201(6) & 0.0119(6) & 0.0368(7) & 0.0060(3) & 0 & 0\\
  \hline
 Si1 & 0.039(1) & $=U_{11}$ & 0.012(1) & 0.0196(6) & 0 & 0\\
 \hline
  \hline
 \multicolumn{7}{|c|}{\textbf{LT-hex}, $Cccm$, $a\times \sqrt3a\times 2c$} \\
 \hline
  \hline
 La1 & 0.0127(6) & 0.025(1) & 0.0177(2) & 0.0011(1) & 0 & 0\\
  \hline
 Ru1/Fe1 & 0.0149(6) & 0.0120(9) & 0.0292(7) & 0 & 0 & 0\\
  \hline
 Ru2 & 0.0122(4) & 0.0213(5) & 0.0190(2) & 0.0041(3) & 0 & 0\\ 
  \hline
 Si1 & 0.021(1) & 0.029(2) & 0.0121(6) & 0.0080(7) & 0 & 0\\
 \hline
  \hline
 \multicolumn{7}{|c|}{\textbf{CO-I}, $a\times \sqrt3a\times 2c$, $Cmmm(0b0)s00~q~=~(0~\frac{1}{2}~0)$} \\
 \hline
  \hline
 La1 & 0.0073(4) & 0.0056(3) & 0.0071(3) & 0 & 0 & 0\\
   $sin$ & 0 & 0 & 0 & 0 & 0 & 0  \\
   $cos$ & 0 & 0 & 0 & -0.0003(2) & 0 & 0\\
  \hline
 La2 & 0.0052(4) & 0.0072(4) & 0.0070(3) & 0 & 0 & 0\\
 $sin$ & 0 & 0 & 0 & 0 & 0 & 0 \\
  $cos$ & 0 & 0 & 0 & 0.0005(2) & 0 & 0\\
  \hline
 Ru1 & 0.0067(4) & 0.0039(3) & 0.0074(2) & 0 & 0 & 0\\
  $sin$ & 0 & 0 & 0 & 0 & 0.0000(1) & 0\\
  $cos$ & 0 & 0 & 0 & 0.0002(2) & 0 & 0\\
  \hline
 Ru2 & 0.0042(2) & 0.0077(2) & 0.0091(1) & -0.0025(2) & 0 & 0\\
  $sin$ & 0 & 0 & 0 & 0 & 0.0001(1) & 0.00068(8)\\
  $cos$ & -0.0002(2) & 0.0002(2) & 0.0018(1) & -0.0000(2) & 0 & 0\\
  \hline
 Si1 & 0.0141(9) & 0.0032(8) & 0.0031(7) & 0 & 0 & 0\\
  $sin$ & 0 & 0 & 0 & -0.0012(7) & 0 & 0\\
  $cos$ & 0 & 0 & 0 & 0.0020(8) & 0 & 0\\
  \hline
 Si1 & 0.0047(9) & 0.0084(9) & 0.0031(7) & 0 & 0 & 0\\
  $sin$ & 0 & 0 & 0 & -0.0010(6) & 0 & 0\\
  $cos$ & 0 & 0 & 0 & 0.0050(6) & 0 & 0\\
 \hline
\end{tabular}
\end{table*}

\newpage


\begin{figure*}[t!]
\centering
\includegraphics[width=1.0\linewidth]{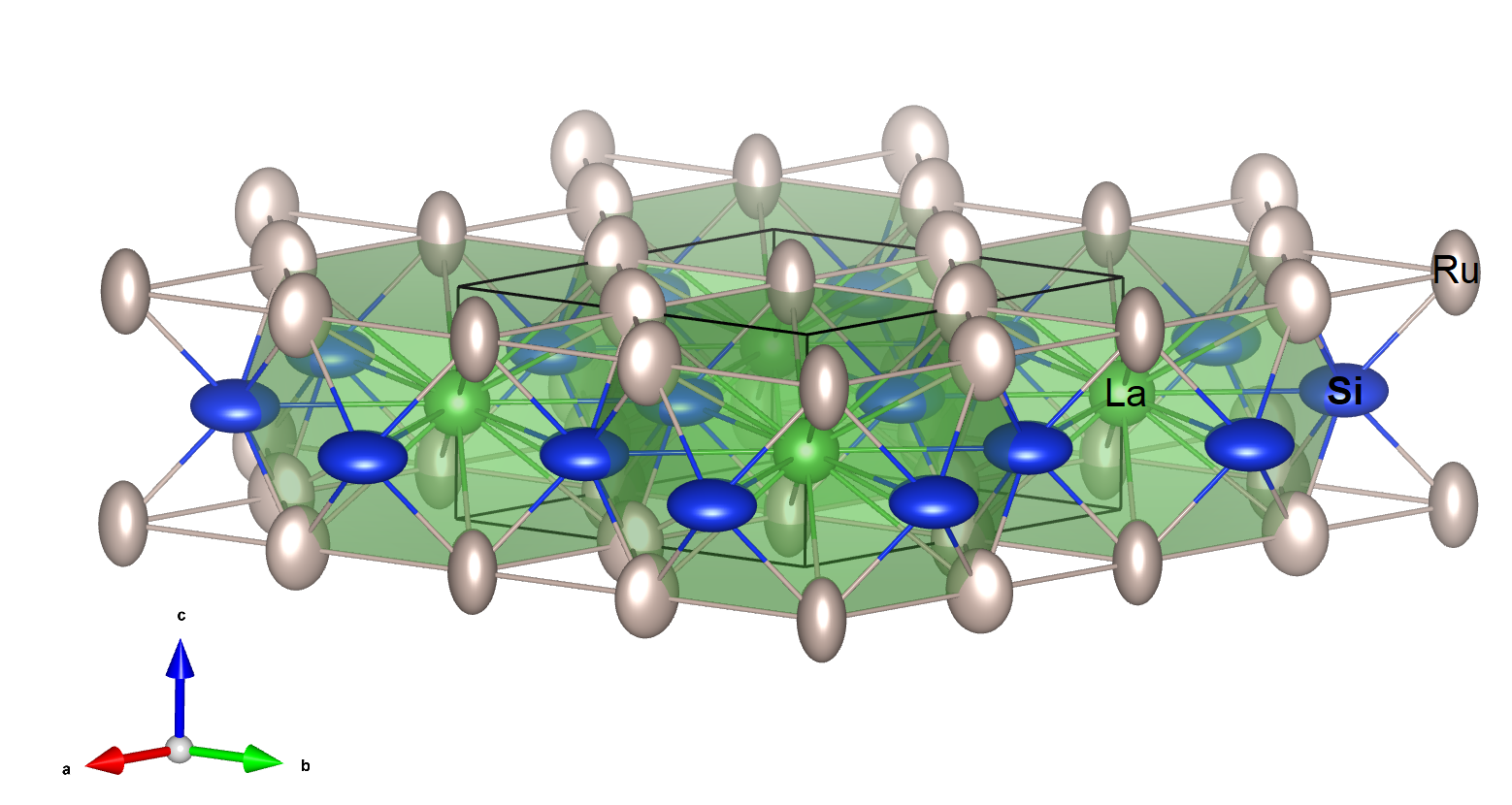}
\vspace{0cm}
\caption{\textbf{High-temperature hexagonal structure.} 
Polyhedral representation of the \textbf{HT-hex} phase. Thermal ellipsoids are drawn at 99${\%}$ probability. The unit cell is outlined by black solid lines.}
\label{fig1}
\end{figure*}

\begin{figure*}[t!]
\centering
\includegraphics[width=1.0\linewidth]{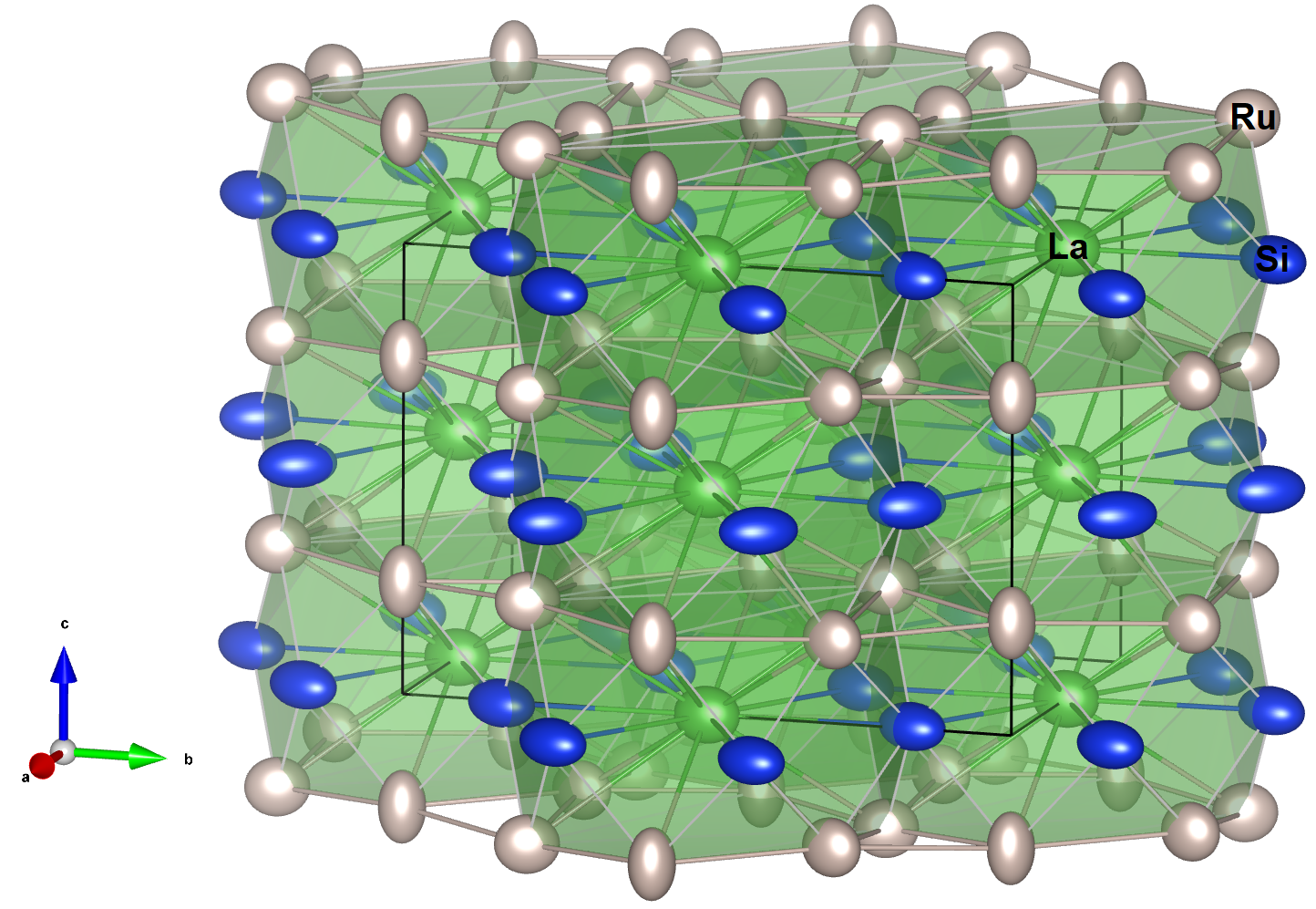}
\vspace{0cm}
\caption{\textbf{Low-temperature hexagonal structure.} 
Polyhedral representation of the \textbf{lt-hex} phase. Thermal ellipsoids are drawn at 99${\%}$ probability. The unit cell is outlined by black solid lines.}
\label{fig1}
\end{figure*}

\begin{figure*}[t!]
\centering
\includegraphics[width=1.0\linewidth]{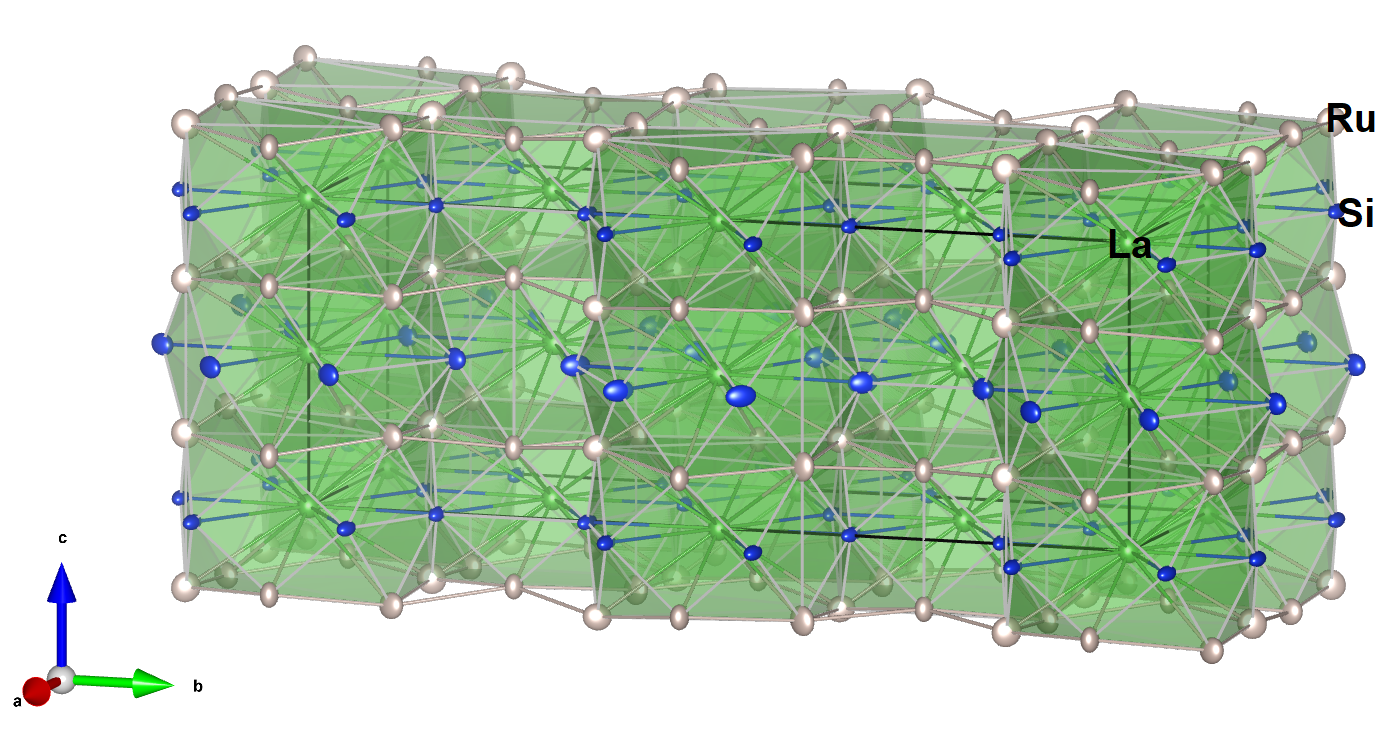}
\vspace{0cm}
\caption{\textbf{Charge order.} 
Polyhedral representation of the CO-I phase. Thermal ellipsoids are drawn at 99${\%}$ probability. The unit cell is outlined by black solid lines. Structures were plotted using the VESTA visualization tool \cite{Vesta}.}
\label{fig1}
\end{figure*}

\begin{figure*}[t!]
\centering
\includegraphics[width=1.0\linewidth]{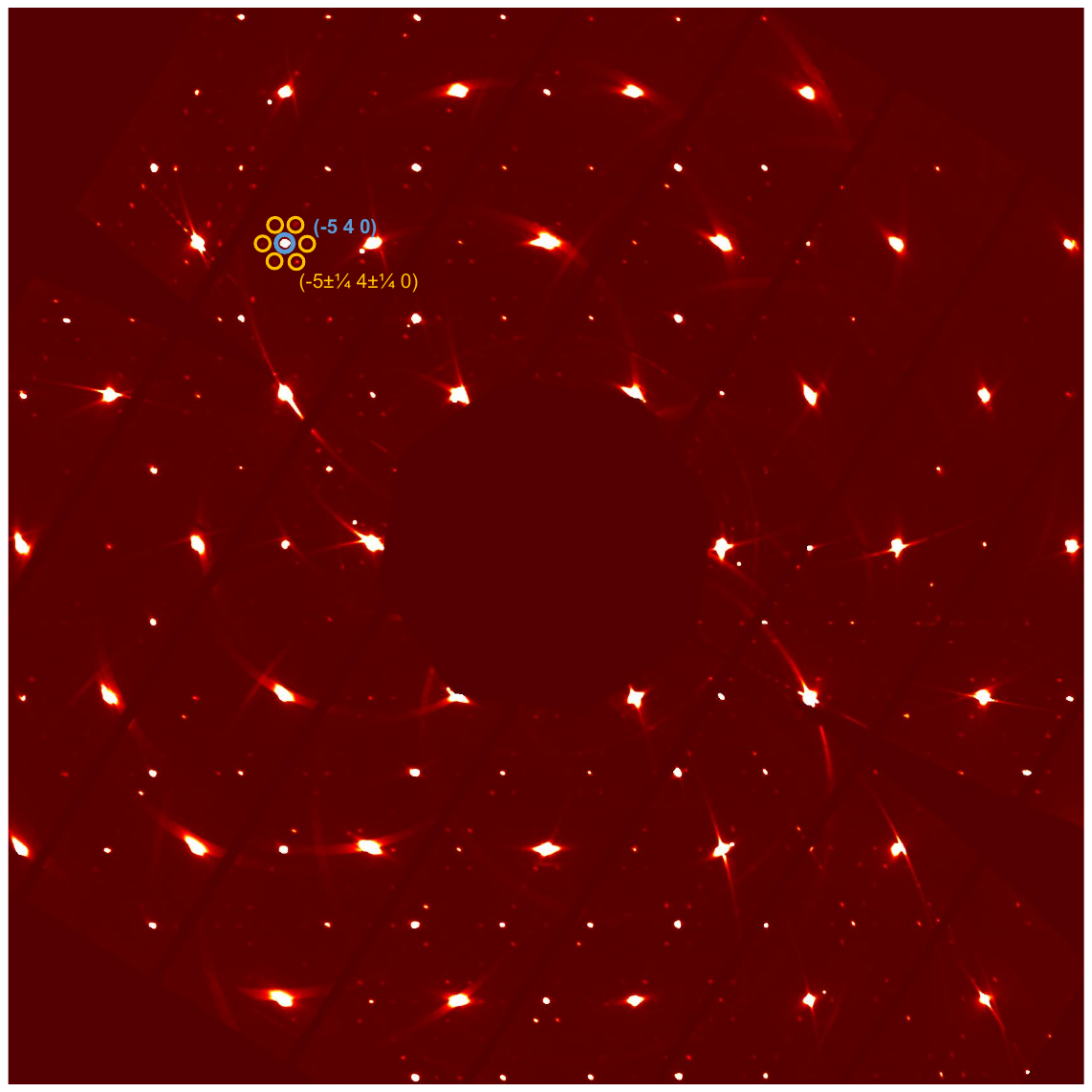}
\vspace{-0.2cm}
\caption{\textbf{Charge order in LaRu$_{3}$Si$_{2}$.} 
Reconstruction of the X-ray diffraction data at 300 K for LaRu$_{3}$Si$_{2}$ collected at the P21.1 beamline at PETRA III using the Pilatus3 X CdTe 2M detector.}
\label{fig1}
\end{figure*}



\end{document}